%% file: main.tex
\documentclass[11pt,a4paper]{article}

\usepackage{jheppub}

\usepackage{amsmath,amssymb}
\usepackage{enumerate}
\usepackage{amsfonts}

\usepackage{subfigure}

\usepackage{allpurpose}

\usepackage{tikz}

\usepackage[marginal,perpage,symbol]{footmisc}

\usepackage{slashed} %

\graphicspath{{./figures/}}

\everymath{\displaystyle}
\makeatletter
\gdef\@fpheader{}
\makeatother

\usepackage[nottoc,notlof,notlot]{tocbibind}

\usepackage{indentfirst}

\begin{document}
\title{Fermionic Love number of \acrlong{rn} black holes}
\date{\today}

\author[a,d,*]{Xiankai Pang,\note{Corresonding author.}}
\emailAdd{xkpang@cwnu.edu.cn}

\affiliation[a]{School of Physics and Astronomy, China West Normal University, Nanchong 637009, China}

\author[b,c]{Yu Tian,}
\affiliation[b]{School of Physical Sciences, University of Chinese Academy of Sciences, Beijing 100049, China}
\affiliation[c]{International of Theoretical Physics, Chinese Academy of Science, Beijing 100190, China}
\emailAdd{ytian@ucas.ac.cn}

\author[d,e]{Hongbao Zhang}
\emailAdd{hongbaozhang@bnu.edu.cn}
\affiliation[d]{School of Physics and Astronomy, Beijing Normal University, Beijing 100875, China}
\affiliation[e]{Key Laboratory of Multiscale Spin Physics, Ministry of Education, Beijing Normal University, Beijing 100875, China}

\author[a]{and Qingquan Jiang}
\emailAdd{qqjiangphys@yeah.net}

\abstract{
	The tidal deformation of compact objects, characterised by their Love numbers, provides insights into the internal structure of neutron stars and black holes. While static bosonic tidal Love numbers vanish for black holes in general relativity, it has been recently revealed that static fermionic tidal perturbations can induce non-zero Love numbers for Kerr black holes. In this paper, we investigate the response of the \acrlong{rn} black hole to the fermionic Weyl field. As a result, we find that the corresponding fermionic tidal Love numbers are also non-vanishing for the \acrlong{rn} black holes except for the extremal ones, which highlights the universal distinct behavior of the static fermionic tidal Love numbers compared to the bosonic counterparts.
}

\keywords{Fermionic Love number; \acrlong{rn} spacetime; Tidal deformation}  

\maketitle

\section{Introduction}
\label{sec:introduction}
The tidal deformability of compact objects, encoded in their \acrlong{tln}s (\acrshort{tln}s), is a fundamental property that quantifies their response to external perturbations \cite{Love:1909yed, Binnington:2009bb, Damour:2009vw, Rodriguez:2026iot, Chakraborty:2026qru}. In particular, \acrshort{tln}s are important probes in gravitational-wave astronomy, as they affect the phase evolution of binary inspiral signals and provide a route to detect the internal structure of compact objects~\cite{Flanagan:2007ix, Hinderer:2007mb}. For example, \acrshort{tln}s of neutron stars are non-zero and depend on the equation of state, allowing for constraints on nuclear matter properties \cite{Damour:2009vw}. In contrast, as a result of the hidden symmetries and ladder structures inherent in the linear bosonic perturbation~\cite{Charalambous:2021kcz,Hui:2021vcv,Hui:2022vbh,BenAchour:2022uqo,Rai:2024lho,Sharma:2024hlz,Combaluzier-Szteinsznaider:2024sgb,Berens:2025lsh,DeLuca:2025zqr,Sharma:2025xii,Cvetic:2026wht,Ghosh:2026vig}, black holes in general relativity exhibit a striking simplicity: their  static bosonic \acrshort{tln}s associated with perturbations by scalar, electromagnetic, and gravitational fields—vanish identically \citep{Binnington:2009bb, Damour:2009vw, Gurlebeck:2015xpa,Poisson:2014gka,Landry:2015zfa,Pani:2015hfa,LeTiec:2020bos,Chia:2020yla,Bhatt:2023zsy}. These null responses align with the no-hair theorem and underscore the unique nature of black holes among astrophysical compact objects~\cite{Gurlebeck:2015xpa}.

A natural and nontrivial question is whether the vanishing of static Love numbers extends beyond bosonic perturbations. From the effective field theory perspective~\cite{Goldberger:2004jt,Porto:2016pyg,Cardoso:2018ptl}, \acrshort{tln}s correspond to response coefficients in a worldline description~\cite{Kol:2011vg,Hui:2020xxx}, and fermionic probes may introduce new operator structures. Moreover, it is also important to test whether the hidden symmetries and ladder structure, which are closely related to the vanishing of static bosonic \acrshort{tln}s, persist in the fermionic sector.
 As a first step towards the answer to this important question, the authors in \cite{Chakraborty:2025zyb} have investigated the static fermionic perturbation of the Kerr black hole. In particular, by defining analogous fermionic TLNs to the bosonic ones, they find that the resulting static fermionic \acrshort{tln}s are non-vanishing, in sharp contrast to the aforementioned bosonic one\footnote{In some scenarios, black holes can also have non-vanishing static bosonic TLNs. In particular, it has been shown very recently that not only does the electric charge lead to non-vanishing \acrshort{tln}s for the charged scalar perturbation on top of the Kerr-Newmann and magnetic black holes~\cite{Ma:2024few,Pereniguez:2025jxq}, but also the non-vanishing cosmological constant, extra spatial dimensions and quantum gravity effect can result in non-zero \acrshort{tln}s for black holes~\cite{Hui:2020xxx,Nair:2022xfm,Nair:2024mya,Franzin:2024cah,Motaharfar:2025typ,Motaharfar:2025ihv,Liu:2025iby,Bhattacharyya:2025slf,Barbosa:2025uau}. In addition, it is worth emphasizing that in the dynamical case, black holes generically exhibit non-zero tidal responses, even for bosonic perturbations~\cite{Chakraborty:2023zed,Chakraborty:2025wvs,Combaluzier-szteinsznaider:2025dtr}.}.
The purpose of this paper is to investigate what happens to the response of the charged \acrfull{rn} black hole to the perturbation by the Weyl neutrino field in the static limit. 
As a result, we find that the corresponding  static \acrshort{tln}s depend on the ratio of the black hole charge to its mass, non-vanishing except in the extremal case $Q=M$, which provides new insights into the tidal properties of black holes.

The paper is organized as follows. In Section~\ref{sec:diracequation}, we first derive the radial equation for the fermionic perturbation in the ingoing Eddington coordinates of the \acrshort{rn} black hole, and then select the regular branch of the resulting exact solutions in the static limit $\omega=0$ by analyzing the asymptotical behavior of the solutions near the horizon. In Section~\ref{sec:lovenumber}, we further expand the resulting regular solution near infinity and read out the corresponding fermionic response, which turns out to be purely real, indicating the generic non-zero fermionic \acrshort{tln}s. We conclude our paper in Section~\ref{sec:discussion} with some discussions. 

Throughout the paper, we choose the metric signature to be $(+---)$.

\section{The Weyl equation in the \acrshort{rn} black hole and its regular static solution}
\label{sec:diracequation}
The \acrshort{rn} black hole can be written as follows
\begin{IEEEeqnarray}{rCl}
	\dd s^2 &=& f(r)\dd t^2-\frac{1}{f(r)}\dd r^2 -r^2\dd\theta^2-r^2\sin^2\theta\dd\varphi^2, ~{ A_{\mu}=\frac{Q}{r}(\dd t)_{\mu}},\label{eq:metricrn}
\end{IEEEeqnarray} 
where and the blackening factor $f(r)$ reads
\begin{IEEEeqnarray}{rCl}
	f(r) &=& 1-\frac{2M}{r}+\frac{Q^2}{r^2}, \label{eq:frrn}
\end{IEEEeqnarray} 
with $M$ and $Q$ the mass and charge of the \acrshort{rn} black hole.  $f(r)=0$ has two roots 
\begin{IEEEeqnarray}{rCl}
	r_+ &=& M+\sqrt{M^2-Q^2}, \quad r_-=M-\sqrt{M^2-Q^2}, \label{eq:defrprm}
\end{IEEEeqnarray} 
corresponding to the outer and inner horizons, respectively. %

When discussing the \acrshort{tln}s, we need to consider the regular solutions to the Weyl equation on the outer horizon. However, the metric~\eqref{eq:metricrn} has the coordinate singularity at $r=r_+$, making it difficult to investigate the regularity of the Weyl neutrino field over there. This obstacle can be well overcome by using the ingoing Eddington coordinate 
\begin{IEEEeqnarray}{rCl}
	v &=& t+r_*,
\end{IEEEeqnarray} 
where $r_*$ is the tortoise coordinate, satisfying $\frac{\dd r_*}{\dd r}=\frac{1}{f(r)}$. 
In terms of such an ingoing Eddington coordinate, the metric~\eqref{eq:metricrn} can be rewritten as 
\begin{IEEEeqnarray}{rCl}
	\dd s^2 &=& f(r)\dd v^2 -2\dd v\dd r-r^2\dd\theta^2-r^2\sin^2\theta\dd\varphi^2,\label{eq:metricrnEddington}
\end{IEEEeqnarray} 
which is obviously regular on the horizon $r=r_+$.

As a normal procedure in the \acrfull{np} formalism (see the appendix~\ref{sec:npformalism} for a brief review), we introduce the null tetrads as follows~\cite{Zhang:2003vb,Ge:2018vjq}
\begin{IEEEeqnarray}{rCl}
	l^\mu &=& \left(1,\frac{f(r)}{2},0,0\right), \IEEEyesnumber \IEEEyessubnumber \label{eq:nulltetradl} \\
	n^\mu &=& \left(0,-1,0,0\right), \IEEEyessubnumber \label{eq:nulltetradn} \\
	m^\mu &=& \frac{1}{\sqrt{2}r}\left(0,0,1,\frac{\ii}{\sin\theta}\right), \IEEEyessubnumber \label{eq:nulltetradm} \\
	\bar{m}^\mu &=& \frac{1}{\sqrt{2}r}\left(0,0,1,-\frac{\ii}{\sin\theta}\right), \IEEEyessubnumber \label{eq:nulltetradmb} 
\end{IEEEeqnarray} 
which are definitely regular on the outer horizon. Consequently, according to the relation~\eqref{eq:eamulnmmb} between the null tetrads and the dyads, the resulting dyads $(o^A,\iota^A)$ are also regular at $r=r_+$. Therefore, the regularity of a spinor field $\psi^A=\psi^1 o^A+\psi^2 \iota^A$ only requires its components $\psi^1$ and $\psi^2$ to be regular on the outer horizon. 

Using Eq.~\eqref{eq:spincoericcicoeaftercal} we can calculate out the non-vanishing spin coefficients as
\begin{IEEEeqnarray}{rCl}
	\varepsilon &=& -\frac{f'(r)}{4}, ~\rho=\frac{f(r)}{2r},  ~\alpha=-\beta=\frac{\cot\theta}{2\sqrt{2}r},~ \mu=\frac{1}{r}. \label{eq:spincoern}
\end{IEEEeqnarray} 
Then substituting these spin coefficients~\eqref{eq:spincoern} into Eqs.~\eqref{eq:diraceqDpsi1} and~\eqref{eq:diraceqdeltapsi1}, we end up with the explicit expression for the Weyl equation 
\begin{equation}
\nabla_{AA'}\psi^A=0
\end{equation}
in the \acrshort{rn} black hole as follows
\begin{IEEEeqnarray}{rCl}
& \left[r f'(r) + 2 f(r) \left(1 + r \frac{\partial }{\partial r}\right) +4r \frac{\partial}{\partial v}\right]\psi^1+ \left[\sqrt{2}\cot\theta +2 \sqrt{2}\left( \frac{\partial }{\partial \theta}  - \frac{\ii}{\sin\theta}\frac{\partial }{\partial \varphi}\right)\right]\psi^2 = 0, &\nonumber \\ 
 \IEEEyesnumber \IEEEyessubnumber \label{eq:diraceqDpsi1rn} \\
& \left[\cot\theta-2\sqrt{2}\left(\frac{\partial}{\partial\theta}+\frac{\ii}{\sin\theta}\frac{\partial}{\partial\varphi}\right)\right]\psi^1 + 4 \left(1+\frac{\partial}{\partial r}\right)\psi^2 = 0.& \IEEEyessubnumber \label{eq:diraceqdeltapsi1rn}
\end{IEEEeqnarray}
Note that our Weyl neutrino field is electrically neutral, so there is no direct interaction between the Weyl neutrino field and the electromagnetic field. Instead, they interact with each other in a indirect way through the gravitational field. 

\subsection{Separation of variables}
\label{sec:varsep}
To solve the equations of motion ~\eqref{eq:diraceqDpsi1rn} and~\eqref{eq:diraceqdeltapsi1rn} in the \acrshort{rn} black hole, we consider the following ansatz
~\cite{Ge:2018vjq,Chakraborty:2025zyb}
\begin{IEEEeqnarray}{rCl}
	\psi^1 &=& \frac{R_-(r)}{r\sqrt{f(r)}}S_-(\theta,\varphi)\ee^{-\ii\omega v}, \quad \psi^2= \frac{R_+(r)}{\sqrt{2}r} S_+(\theta,\varphi)\ee^{-\ii\omega v},  \label{eq:separationansatz}
\end{IEEEeqnarray} 
where $S_{\pm}(\theta,\varphi)={}_{\pm \frac{1}{2}} Y_{lm}(\theta,\varphi)$ is the spin-weighted spherical harmonics with $s=\pm \frac{1}{2}$~\cite{Goldberg:1966uu,Penrose:1985bww}. For convenience, the explicit form and relevant properties of ${}_s Y_{lm}$ are given in the appendix~\ref{sec:swsh}. 

Substituting the ansatz~\eqref{eq:separationansatz} into our equations of motion~\eqref{eq:diraceqDpsi1rn} and~\eqref{eq:diraceqdeltapsi1rn}, we obtain the following coupled radial equations for $R_-$ and $R_+$
\begin{IEEEeqnarray}{rCl}
	\left(-2\ii\omega +\frac{\dd}{\dd r_*}\right)R_-(r) &=& -\lambda_1\frac{\sqrt{f(r)}}{r}R_+(r),\IEEEyesnumber \IEEEyessubnumber \label{eq:diraceqradialRm}\\
	\quad \frac{\dd}{\dd r_*}R_+(r)&=&-\lambda_2 \frac{\sqrt{f(r)}}{r}R_-(r),\IEEEyessubnumber \label{eq:diraceqradialRp}
\end{IEEEeqnarray} 
where $\lambda_1$ and $\lambda_2$ are two separation constants, to be fixed through the following angular equations~\cite{Goldberg:1966uu,Penrose:1985bww}
\begin{IEEEeqnarray*}{rCl}
	\bar{\eth} S_+ &=&-\left(\partial_\theta -\frac{\ii}{\sin\theta}\partial_{\varphi}+\frac{1}{2}\cot\theta\right)S_+= -\lambda_1 S_-,  \\
	\eth S_- &=& -\left(\partial_\theta+\frac{\ii}{\sin\theta}\partial_\varphi+\frac{1}{2}\cot\theta\right)S_-= \lambda_2 S_+.
\end{IEEEeqnarray*} 
Comparing with Eqs.~\eqref{eq:ethsYlm} and~\eqref{eq:ethbsYlm} with $s=\pm \frac{1}{2}$, we obtain
\begin{IEEEeqnarray*}{rCl}
	\lambda_1 &=& \sqrt{l(l+1)-\frac{1}{2}\left(\frac{1}{2}-1\right)} = l+\frac{1}{2},  \\
	\lambda_2 &=& \sqrt{l(l+1)-\left(-\frac{1}{2}\right)\left[\left(-\frac{1}{2}\right)-1\right]} = l+\frac{1}{2} \\
\end{IEEEeqnarray*} 
with the half integer $l\geq\frac{1}{2}$, which implies that 
\begin{IEEEeqnarray}{rCl}
	\lambda_1=\lambda_2=\lambda&\equiv&l+\frac{1}{2}\label{eq:lambdalrelation}
\end{IEEEeqnarray} 
with the integer $\lambda \geq 1$.

\subsection{Selection of the regular static solution to the radial equations}
\label{sec:staticsolution}
To solve the radial Eqs.~\eqref{eq:diraceqradialRm} and~\eqref{eq:diraceqradialRp}, we can decouple $R_-$ and $R_+$ by taking a further derivative with respect to the radius $r$ on both sides of Eq.~\eqref{eq:diraceqradialRm} and eliminating $\dd R_+/\dd r$ using Eq.~\eqref{eq:diraceqradialRp}. This will result in a second-order equation for $R_-$
\begin{align}
& z (z+1) \Big( 2 z (z+1) (r_+-r_-) R_-''(z)  +  \left\{ (2 z+1) (r_+-r_-) - 4 i \omega [r_++(r_+-r_-)z]^2 \right\}R_-'(z) \Big) \nonumber \\
& - 2 \Big\{ \lambda ^2 z (z+1) (r_+-r_-) + i \omega [r_++(r_+-r_-)z] \left[ r_+ (2 z^2 + z - 1) - r_- z (2 z + 3) \right] \Big\}R_-(z)  = 0. \label{eq:Rmzppomega}
\end{align}
Here the prime denotes the derivative with respect to the dimensionless variable $z$, which is defined as follows%
\begin{IEEEeqnarray}{rCl}
	z &=& \frac{r-r_+}{r_+-r_-}. \label{eq:defz}
\end{IEEEeqnarray} 
We see that the dimensionless variable $z\rightarrow 0$ as the radius $r\rightarrow r_+$, and $z\rightarrow\infty$ as  $r\to\infty$.

For our present purpose, we shall focus on the static solution to Eq.~\eqref{eq:Rmzppomega}, which reduces in the static limit $\omega=0$ to the following much simpler form
\begin{IEEEeqnarray}{rCl}
	2z(1+z)R_-''(z)+(1+2z)R_-'(z)-2\lambda^2R_-(z) &=&0  .\label{eq:Rmzpp}
\end{IEEEeqnarray} 
Whence one can obtain the corresponding exact solution
\begin{IEEEeqnarray}{rCl}
	R_- &=& c_1 \cosh \left(2\lambda \sinh^{-1}\sqrt{z}\right)+\frac{c_2\sinh \left(2\lambda \sinh^{-1}\sqrt{z}\right)}{2\lambda} .  \label{eq:Rmsol}
\end{IEEEeqnarray} 
As alluded to before, the regularity amounts to requiring that 
both $R_- / \sqrt{f}$  and $R_+$ should remain finite on the outer horizon. Note that our blackening factor $f$ in terms of $z$ reads
\begin{IEEEeqnarray}{rCl}
	f(z) &=& \frac{(r_+-r_-)^2z(1+z)}{[r_++(r_+-r_-)z]^2},
\end{IEEEeqnarray} 
whose asymptotic behavior goes as
\begin{IEEEeqnarray}{rCl}
	f(z) &\to& \frac{(r_+-r_-)^2z}{r_+^2}, \label{eq:fz0exp}
\end{IEEEeqnarray} 
as one approaches the outer horizon $z=0$.

Accordingly, as $z\rightarrow 0$,  $R_-/\sqrt{f}$ has the following asymptotic behavior
\begin{IEEEeqnarray*}{rCl}
	\frac{R_-}{\sqrt{f}} &\to& \frac{r_+}{(r_+ - r_-) \sqrt{z}} c_1 + \frac{r_+}{ r_+ - r_-} c_2 .
\end{IEEEeqnarray*}
To ensure the regularity at the outer horizon, we must set $c_1 = 0$. In addition, as explained later on,
the \acrshort{tln}s are independent of the overall factor $c_2$. So for simplicity but without loss of generality, below we set $c_2 = 1$, which means that our regular solution $R_-$ reduces to the following form
\begin{IEEEeqnarray}{rCl}
	R_-(z) &=& \frac{\sinh \left(2\lambda \sinh^{-1}\sqrt{z}\right)}{2\lambda}. \label{eq:Rmsolreg}
\end{IEEEeqnarray}
The corresponding solution for $R_+$ can be derived from Eq.~\eqref{eq:diraceqradialRm} in the static limit $\omega = 0$, reading
\begin{IEEEeqnarray}{rCl}
	R_+(z) &=& -\frac{\cosh \left(2\lambda \sinh^{-1}\sqrt{z}\right)}{2\lambda}. \label{eq:Rpsolreg}
\end{IEEEeqnarray}
which also behaves regularly as
\begin{IEEEeqnarray*}{rCl}
	R_+(z) &\to& -\frac{1}{2\lambda}
\end{IEEEeqnarray*}
near the outer horizon. Thus we have accomplished the selection of the regular static solution to the Weyl equation in the \acrshort{rn} black hole.

\section{The fermionic \acrshort{tln}s of the \acrshort{rn} black hole}
\label{sec:lovenumber}
The response of an \acrshort{rn} black hole to the fermionic Weyl perturbations can be extracted from the asymptotic expansion of the previously obtained static radial solution at infinity.
As such, we like to resort to the following mathematical relation~\cite{Gradshteyn:1996table}
\begin{IEEEeqnarray*}{rCl}
	{}_2F_1 \left(\frac{1}{2}-\lambda,\frac{1}{2}+\lambda,\frac{3}{2},-z\right) &=& \frac{\sinh \left(2\lambda\sinh^{-1}\sqrt{z}\right)}{2\lambda\sqrt{z}}
\end{IEEEeqnarray*} 
with $_2F_1$ the hypergeometric function. Accordingly, our regular solution~\eqref{eq:Rmsolreg} can be rewritten as follows
\begin{IEEEeqnarray}{rCl}
	R_-(z) &=& \sqrt{z} {}_2F_1 \left(\frac{1}{2}-\lambda,\frac{1}{2}+\lambda,\frac{3}{2},-z\right),\label{eq:Rm2F1}
\end{IEEEeqnarray} 
which has the following asymptotic behavior as $z\rightarrow\infty$
\begin{align}
	R_-  &\to\frac{\Gamma\left(-\frac{1}{2}\right)\Gamma(2l+1)}{\Gamma(l+1)\Gamma\left(l+\frac{3}{2}\right)} z^{l+\frac{1}{2}}\left[(1+\cdots)+\frac{\Gamma\left(l+\frac{3}{2}\right)\Gamma\left(-l-\frac{1}{2}\right)}{4^{2l+1}\Gamma\left(l+\frac{1}{2}\right)\Gamma\left(-l+\frac{1}{2}\right)}z^{-(2l+1)}(1+\cdots)\right],
    \nonumber \\ 
		&\to -\frac{4^{l+1}}{2l+1}  \frac{r^{l+\frac{1}{2}}}{(r_+-r_-)^{l+\frac{1}{2}}} \left[(1+\cdots)-\frac{1}{4^{2l+1}}\left(1-\frac{r_-}{r_+}\right)^{2l+1}\left(\frac{r_+}{r}\right)^{2l+1}(1+\cdots)\right].\label{eq:Rmsolrnexp}
\end{align}
Here we have used the following identities of the $\Gamma$ function~\cite{Gradshteyn:1996table} %
\begin{IEEEeqnarray*}{rCl}
	\frac{\Gamma\left(-\frac{1}{2}\right)\Gamma(2l+1)}{\Gamma(l+1)\Gamma\left(l+\frac{3}{2}\right)}&=&-\frac{4^{l+1}}{2l+1},\quad\frac{\Gamma \left(l+\frac{3}{2}\right)}{\Gamma \left(l+\frac{1}{2}\right)} =-\frac{\Gamma \left(-l+\frac{1}{2}\right)}{\Gamma \left(-l-\frac{1}{2}\right)} = l+\frac{1}{2} ,
\end{IEEEeqnarray*} 
with the dot denoting the sub-leading terms, which are irrelevant to our discussion. In particular, the corresponding response is defined as the coefficient of $\left(\frac{r_+}{r}\right)^{2l+1}$ in the square bracket~\cite{Chakrabarti:2013xza,Chia:2020yla,Chakraborty:2025zyb}, which reads
\begin{IEEEeqnarray}{rCl}
	\mathcal{F}_{-\frac{1}{2}lm} &=& -\frac{1}{4^{2l+1}}\left(1-\frac{r_-}{r_+}\right)^{2l+1},
\end{IEEEeqnarray} 
independent of the azimuthal number $m$ due to the spherical symmetry of the \acrshort{rn} black hole.%

Following the same procedure for $R_+$, we can extract the resulting response $\mathcal{F}_{\frac{1}{2}lm}$, which is related to $\mathcal{F}_{-\frac{1}{2}lm}$ as $\mathcal{F}_{\frac{1}{2}lm}=-\mathcal{F}_{-\frac{1}{2}lm}$. Using the dimensionless parameter $q=Q/M$, we can express the above responses of the \acrshort{rn} black hole to the fermionic Weyl field perturbations as 
\begin{IEEEeqnarray}{rCl}
	\mathcal{F}_{\pm \frac{1}{2}lm} &=& \pm \left[\frac{\sqrt{1-q^2}}{2 \left(1+\sqrt{1-q^2}\right)}\right]^{2l+1} , \label{eq:F1p2lmq}
\end{IEEEeqnarray} 
where we have used Eq.~\eqref{eq:defrprm} for $r_+$ and $r_-$. 

When $q=0$, we reproduce the corresponding result for the Schwarzschild black hole \cite{Chakraborty:2025zyb}
\begin{IEEEeqnarray}{rCl}
  \mathcal{F}_{\pm \frac{1}{2}lm}^{\text{Sch}}&=&\pm\frac{1}{4^{2l+1}}, 
\end{IEEEeqnarray}
{ whereby we can rewrite Eq. \eqref{eq:F1p2lmq} as follows}
\begin{IEEEeqnarray}{rCl}
    \mathcal{F}_{\pm \frac{1}{2}lm}& =& \mathcal{F}_{\pm \frac{1}{2}lm}^{\text{Sch}}\left(\frac{2\sqrt{1-q^2}}{1+\sqrt{1-q^2}}\right)^{2l+1}. 
\end{IEEEeqnarray}
Accordingly, as the black hole under consideration is changed from the Schwarzschild black hole to the extremal one, namely $q$ is increased from $0$ to $1$\footnote{As shown in Appendix C, Eq.~\eqref{eq:F1p2lmq} also applies to the extremal \acrshort{rn} black hole although the definition of the dimensionless variable $z$ breaks down.}, $\mathcal{F}_{\pm \frac{1}{2}lm}$ deviates from $\mathcal{F}^\text{Sch}_{\pm \frac{1}{2}lm}$ larger and larger, becoming exactly zero for the extremal black hole. On the other hand, for $0<q<1$, the larger $l$ is, the closer to zero the ratio of $\mathcal{F}_{\pm \frac{1}{2}lm}$ to $\mathcal{F}^\text{Sch}_{\pm \frac{1}{2}lm}$ is.

Note that the \acrshort{rn} black hole solution requires $0\leq q\leq 1$. As a result, the fermionic responses~\eqref{eq:F1p2lmq} are purely real, indicating that
\begin{itemize}
    \item the corresponding static fermionic \acrshort{tln}s, proportional to the real part of $\mathcal{F}_{\pm \frac{1}{2}lm}$~\cite{Chakrabarti:2013xza,Chakraborty:2025zyb}, are non-vanishing except in the case of the extremal \acrshort{rn} black hole with $q=1$;
    \item the corresponding static fermionic dissipation numbers~\cite{Goldberger:2005cd}, proportional to the imaginary part of $\mathcal{F}_{\pm \frac{1}{2}lm}$, are exactly zero, the same as the Kerr black hole result~\cite{Chakraborty:2025zyb}.
\end{itemize}

 It is instructive to contrast our result with the bosonic case. For neutral scalar, electromagnetic, and gravitational perturbations, the static \acrshort{tln}s of black holes vanish identically. In contrast, the fermionic perturbations lead to non-vanishing static response coefficients (except for extremal \acrshort{rn} black holes). This highlights a qualitative distinction between fermionic and bosonic probes of black hole spacetimes, suggesting that the underlying symmetry responsible for vanishing bosonic \acrshort{tln}s does not straightforwardly extend to the fermions.

\section{Conclusions}
\label{sec:discussion}

In this paper we have derived, for the first time, the explicit expression for the static response of the \acrshort{rn} black hole to the fermionic Weyl field, which turns out to be purely real, in contrast to the purely imaginary static response to the neutral bosonic fields. As a result, the static fermionic \acrshort{tln}s of the \acrshort{rn} black hole are non-zero except for the extremal case. Along with \cite{Chakraborty:2025zyb}, our work further underscores the importance of considering fermionic fields in probing black hole properties. 

There are also various generalizations worthy of further investigation. First, the response of the black hole to the bosonic fields depends on the spacetime dimension~\cite{Rodriguez:2023xjd}. So it is intriguing to see what happens to the response of the higher-dimensional \acrshort{rn} black hole to the fermionic fields. Second, our Weyl neutrino field is uncharged. But as mentioned in the introduction section, the response of the black hole to the charged scalar field demonstrates a distinct behavior. So it is also interesting to consider the response of the \acrshort{rn} black hole to the charged fermionic Dirac field. Last but not least, so far what we have considered is the static response, it is also important for us to explore the dynamic response of the \acrshort{rn} black hole to fermionic fields, although one is required to resort to numerics because one may have no exact solutions any more for the dynamic case. We will report these potential generalizations somewhere else in the future.

\acknowledgments

This work is partially supported by the National Key Research and Development Program of China with Grant No. 2021YFC2203001 as well as the National Natural Science Foundation of China (NSFC) with Grant Nos. 12035016, 12275350, 12375048, 12375058, 12361141825, 12447182, 12575047, and 12505082. 
XP is also supported by the Doctoral Initiation Grant 24KE051 and  Basic Research Grant 25kx010 from China West Normal University. QJ is supported by the Key Joint Program of Science and Education of Sichuan Province with Grant No. 25LHJJ0097.

\appendix 
\section{The \acrlong{np} formalism}
\label{sec:npformalism}
In this appendix we will provide a brief review of the  \acrshort{np} formalism as well as its relation with the spinors, whereby we derive the explicit expression for the Weyl equation in curved spacetime. For more details, please refer to \cite{Wald:1984rg,Chandrasekhar:1985kt,Penrose:1985bww}. 
\paragraph{The dyads.} Every $2$-spinor can be projected onto the spinor basis $(o^A,\iota^A)$
\begin{IEEEeqnarray}{rCl}
	\psi^A &=& \psi^1 o^A+\psi^2 \iota^A . 
\end{IEEEeqnarray} 
For simplicity, we can use the dyads $(\xi_{\Sigma})^A=(o^A,\iota^A)$ denote the basis
\begin{IEEEeqnarray}{rCl}
	\left(\xi_0\right)^A&=&  o^A, ~ \left(\xi_1\right)^A=\iota^A  . \label{eq:defdyadxi0xi1}
\end{IEEEeqnarray} 

We choose the dyads such that $(o^A,\iota^A)$ span the $\left(\frac{1}{2},0\right)$ represent space of the Lorentz group~\cite{Weinberg:1995mt,Schwartz:2013pla}.  Their complex conjugate $(o^{A'},\iota^{A'})$ then span the $\left(0,\frac{1}{2}\right)$, and the corresponding indices are decorated by a prime.

\paragraph{The skew metric.} In the calculations involving spinor, it's very useful to introduce the skew metric~\cite{Chandrasekhar:1985kt} 
\begin{IEEEeqnarray}{rCl}
	\epsilon^{AB} &=& o^A\iota^B-\iota^A o^B , \label{eq:epsilonuuAB}
\end{IEEEeqnarray} 
which is antisymmetric with respect to $A$ and $B$. The one with lowered indices can be given as the inverse of Eq.~\eqref{eq:epsilonuuAB}, such that~\cite{Chandrasekhar:1985kt}
\begin{IEEEeqnarray}{rCl}
	\epsilon_{CA}\epsilon^{CB} &=& \delta^B_A . \label{eq:epsilonCACBcontract}
\end{IEEEeqnarray} 

We can use $\epsilon_{AB}$ to lower the indices of the spinor basis as follows 
\begin{IEEEeqnarray}{rCl}
	o_A &=& \epsilon_{BA}o^B ,~  \iota_A=\epsilon_{BA}\iota^B , \label{eq:ouAiotauAlowering}
\end{IEEEeqnarray} 
while the raising of the indices is given by\footnote{Note that the ordering of the indices is very important.}
\begin{IEEEeqnarray}{rCl}
	o^A &=& \epsilon^{AB}o_B, ~\iota^A=\epsilon^{AB}\iota_B. 
\end{IEEEeqnarray} 

Since $\epsilon_{AB}$ is antisymmetric, we see that 
\begin{IEEEeqnarray}{rCl}
	o_A o^A &=& \iota_A\iota^A=0 . \label{eq:ooiotaiotacontract}
\end{IEEEeqnarray} 
Furthermore, if the dyads are normalised such that 
\begin{IEEEeqnarray}{rCl}
	o_A\iota^A &=& -o^A\iota_A=1, \label{eq:oiotanorm}
\end{IEEEeqnarray} 
the skew metric would be the two-dimensional Levi-Civita symbol~\cite{Chandrasekhar:1985kt,Penrose:1985bww} 
\begin{IEEEeqnarray}{rCl}
	\epsilon^{\Sigma\Lambda} &=&\epsilon_{\Sigma\Lambda}=\left(\begin{array}{cc}
			0 & 1 \\
			-1 & 0
	\end{array}\right) , \label{eq:epsiloncompmat}
\end{IEEEeqnarray} 
which is invariant under the action of $SL(2,C)$ group. 

\paragraph{The null tetrads.} It's well known that every null vector in $4$d spacetime can be written as a product of a spinor~\cite{Wald:1984rg,Penrose:1985bww}.
And for simplicity we write the null tetrads $e_a^{~\mu}=(l^\mu,n^\mu,m^\mu,\bar{m}^\mu)$ associated to the dyads~\eqref{eq:defdyadxi0xi1} as following~\cite{Chandrasekhar:1985kt}
\begin{IEEEeqnarray}{rCl}
	e_1^{~\mu}&=& l^\mu\sim o^A\bar{o}^{A'} , ~ e_2^{~\mu}= n^\mu\sim\iota^A\bar{\iota}^{A'} , ~ e_3^{~\mu}= m^\mu\sim o^A\bar{\iota}^{A'} , ~ e_4^{~\mu}=\bar{m}^\mu\sim \iota^A\bar{o}^{A'} , \label{eq:eamulnmmb}
\end{IEEEeqnarray} 
whose coordinate index $\mu$ can be lowered using the spacetime metric $g_{\mu\nu}$ as usual 
\begin{IEEEeqnarray}{rCl}
	e_{a\mu} &=& g_{\mu\nu} e_a^{~\nu}.
\end{IEEEeqnarray} 
And the $\sim$ symbol in Eq.~\eqref{eq:eamulnmmb} indicates that the null tetrads can be obtained from the products of dyads. For example, 
\begin{IEEEeqnarray}{rCl}
	l^\mu&=&\sigma^\mu_{AA'}o^A\bar{o}^{A'} , 
\end{IEEEeqnarray} 
where the matrices $\sigma^\mu$ are defined as~\cite{Penrose:1985bww,Futterman:1988ni}
\begin{IEEEeqnarray}{rCl}
	\sigma^\mu &=& \frac{1}{\sqrt{2}}\left(I, \sigma^i\right) ,
\end{IEEEeqnarray}
for the $2\times 2$ identity matrix $I$ and the Pauli matrices $\sigma^i$.

Using the Eqs.~~\eqref{eq:ooiotaiotacontract} and~\eqref{eq:oiotanorm}, 
we can easily get the null metric
\begin{IEEEeqnarray}{rCl}
	\eta_{ab} &\equiv& e_a^{~\mu}e_{b\mu}= \left(\begin{array}{cc}
			\sigma^1 & 0 \\
			0 & -\sigma^1
	\end{array}\right) . \label{eq:nullmetric} 
\end{IEEEeqnarray} 

In curved spacetimes, the null tetrads~\eqref{eq:eamulnmmb} are coordinate dependent, whose behavior under covariant derivative can be measured by the \emph{Ricci rotation coefficients}~\cite{Wald:1984rg}. In terms of components, these coefficients read~\cite{Pang:2024tco} 
\begin{IEEEeqnarray}{rCl}
	\Gamma_{abc} &=& (e_{a\mu})_{;\nu}e_b^{~\mu}e_c^{\nu}. \label{eq:Gammaldldldabc}
\end{IEEEeqnarray} 

Finally, we denote the directional derivative along the null tetrads as follows~\cite {Penrose:1985bww}
\begin{IEEEeqnarray}{rCl}
	D &=& o^A\bar{o}^{A'}\nabla_{AA'}=l^\mu\nabla_{\mu} , \IEEEyessubnumber \label{eq:defDoAoAp} \\
	\Delta &=&  \iota^A\bar{\iota}^{A'}\nabla_{AA'}=n^\mu\nabla_{\mu} ,\IEEEyessubnumber \label{eq:defDeltaiotaAiotaAp} \\ 
	\delta &=& o^A\bar{\iota}^{A'}\nabla_{AA'}=m^\mu\nabla_{\mu}, \IEEEyessubnumber \label{eq:defdeltaoAiotaAp} \\ 
	\bar{\delta} &=& \iota^A\bar{o}^{A'}\nabla_{AA'}=\bar{m}^\mu\nabla_{\mu} . \IEEEyessubnumber \label{eq:defdeltabiotaAoAp}
\end{IEEEeqnarray}

\paragraph{The spin coefficients.} %
The covariant derivative of the dyads can be given by the \emph{spin coefficients}~\cite{Penrose:1985bww}
\begin{IEEEeqnarray}{rCl}
	\Gamma_{AA'\Sigma\Lambda} &=& \left(\xi_\Sigma\right)_B\nabla_{AA'}\left(\xi_{\Lambda}\right)^B  . \label{eq:defGammaxi}
\end{IEEEeqnarray} 
Or, in terms of components~\cite{Wald:1984rg}
\begin{IEEEeqnarray}{rCl}
	\Gamma_{\Psi\Psi'\Sigma\Lambda} &=&  \left(\xi_\Psi\right)^A \left(\bar{\xi}_{\Psi'}\right)^{A'}\left(\xi_\Sigma\right)_B\nabla_{AA'}\left(\xi_{\Lambda}\right)^B. \label{eq:defGammacompxi}
\end{IEEEeqnarray} 
One can easily show that the spin coefficients are symmetric with respect to the exchange of $\Sigma$ and $\Lambda$, and therefore, the spin coefficients $\Gamma_{\Psi\Psi'\Sigma\Lambda}$ have only $12$ independent components.

The spin coefficients $\Gamma_{\Psi\Psi'\Sigma\Lambda}$ can be calculated using Ricci rotation coefficients $\Gamma_{abc}$ thanks to the following identity~\cite{Wald:1984rg}
\begin{IEEEeqnarray}{rCl}
	\Gamma_{\Psi\Psi'\Sigma\Lambda} &=& (\xi_{\Psi})^A \left(\bar{\xi}_{\Psi'}\right)^{A'}(\xi_{\Sigma})_B\nabla_{AA'}(\xi_{\Lambda})^B \quad, \nonumber\\
									&=&\frac{1}{2} (\xi_{\Psi})^A \left(\bar{\xi}_{\Psi'}\right)^{A'}\bar{\epsilon}^{\Gamma'\Delta'}(\bar{\xi}_{\Gamma'})_{B'}(\xi_{\Sigma})_B\nabla_{AA'}\left[(\xi_{\Lambda})^B(\bar{\xi}_{\Delta'})^{B'}\right] .  \label{eq:GammaspinRicciidentitycomp}
\end{IEEEeqnarray} 
For example, 
using Eq.~\eqref{eq:GammaspinRicciidentitycomp} for $\Gamma_{00'00}$ one obtains
\begin{IEEEeqnarray*}{rCl}
	\Gamma_{00'00} &=& \frac{1}{2}o^A\bar{o}^{A'}\bar{\epsilon}^{\Gamma'\Delta'}(\bar{\xi}_{\Gamma'})_{B'}o_B\nabla_{AA'}\left[o^B \left(\bar{\xi}_{\Delta'}\right)^{B'}\right] , \\
				   &=& \frac{1}{2} o^A\bar{o}^{A'}o_B \left[\bar{o}_{B'}\nabla_{AA'}\left(o^B\bar{\iota}^{B'}\right)-\bar{\iota}_{B'}\nabla_{AA'}\left(o^B\bar{o}^{B'}\right)\right] ,  \\
				   &=& \frac{1}{2}\left(l_{\mu}l^\nu\nabla_{\nu}m^\mu-m_{\mu}l^{\nu}\nabla_{\nu}l^{\mu}\right),   \\ 
				   &=& (e_{3\mu})_{;\nu}e_1^{~\mu}e_1^{~\nu} ,  \\ 
				   &=& \Gamma_{311} ,  
\end{IEEEeqnarray*} 
where we have used the definition~\eqref{eq:eamulnmmb} of the null tetrads $e_a^{~\mu}$, and $\Gamma_{abc}$ denotes the corresponding Ricci rotation coefficient given by Eq.~\eqref{eq:Gammaldldldabc}.

It is customary to represent the components of spin coefficients using certain symbols, and Eq.~\eqref{eq:spincoericcicoeaftercal} shows the definition of these symbols~\cite{Penrose:1985bww} and their relation to Ricci rotation coefficients~\eqref{eq:Gammaldldldabc}
\begin{equation}
	\Gamma_{\Psi\Psi'\Sigma\Lambda} = 
\begin{NiceTabular}{|cc|c|c|c|}
\toprule
\Block{2-2}{\diagbox{$\Psi \Psi'$}{$\Sigma\Lambda$}}  & & \Block{2-1}{$00$} & $10$ & \Block{2-1}{$11$} \\
														& & &or $01$ &\\
														\toprule %
														&  $00'$ & $\kappa=\Gamma_{311}$ & $\varepsilon=\dfrac{1}{2}\left( \Gamma_{211} + \Gamma_{341} \right)$ & $\pi=\Gamma_{241} $ \\
\midrule
														&  $10'$ & $\rho= \Gamma_{314}$ & $\alpha= \dfrac{1}{2}\left( \Gamma_{214} + \Gamma_{344} \right)$ & $\lambda= \Gamma_{244}$ \\
\midrule
														& $01'$ & $\sigma= \Gamma_{313}$ & $\beta= \dfrac{1}{2}\left( \Gamma_{213} + \Gamma_{343} \right) $ & $\mu= \Gamma_{243} $ \\
\midrule
														&  $11'$ & $\tau= \Gamma_{312}$ & $\gamma= \dfrac{1}{2}\left( \Gamma_{212} + \Gamma_{342} \right) $ & $\nu= \Gamma_{242}  $ \\
\bottomrule
\end{NiceTabular} \label{eq:spincoericcicoeaftercal}.
\end{equation}

\paragraph{The Weyl equation.} Finally, we can write down the Weyl equation in the \acrshort{np} formalism
\begin{IEEEeqnarray}{rCl}
	\nabla_{AA'}\psi^A &=&0  .
\end{IEEEeqnarray} 
In terms of components, we have 
\begin{IEEEeqnarray}{rCl}
	\bar{o}^{A'}\nabla_{AA'} \psi^A&=&\bar{o}^{A'}\nabla_{AA'}\left(\psi^1 o^A+\psi^2 \iota^A\right)  ,  \nonumber \\
					   &=& (D\psi^1)+\psi^1\bar{o}^{A'}\nabla_{AA'}o^A+(\bar{\delta} \psi^2)+\psi^2\bar{o}^{A'}\nabla_{AA'}\iota^A, \\ 
			\bar{\iota}^{A'}\nabla_{AA'}\psi^A &=& \bar{\iota}^{A'}\nabla_{AA'}\left(\psi^1 o^A+\psi^2 \iota^A\right)  ,  \nonumber \\ 
											   &=& (\delta\psi^1)+\psi^1\bar{\iota}^{A'}\nabla_{AA'}o^A+(\Delta \psi^2)+\psi^2\bar{\iota}^{A'}\nabla_{AA'}\iota^A  ,  
\end{IEEEeqnarray} 
where we have used the definitions~\eqref{eq:defDoAoAp}-\eqref{eq:defdeltabiotaAoAp} of the directional derivatives\footnote{When acting on components, for example $\psi^1$, the covariant derivative $\nabla_\mu=\partial_\mu$ is just the ordinary derivative, therefore $D\psi^1=l^\mu\partial_\mu \psi^1$.}.

Note that terms like $\bar{o}^{A'}\nabla_{AA'}o^A$ can be calculated using the following identity
\begin{IEEEeqnarray}{rCl}
	\left(\bar{\xi}_{\Delta'}\right)^{A'}\nabla_{AA'}\left(\xi_{\Psi}\right)^A &=&\epsilon^{\Lambda\Delta}\Gamma_{\Delta\Delta'\Lambda\Psi}  , \label{eq:nablaGammarelation} 
\end{IEEEeqnarray} 
which can be proved by direct calculations. Using the identity~\eqref{eq:nablaGammarelation} we obtain
\begin{IEEEeqnarray*}{rCl}
	\bar{o}^{A'}\nabla_{AA'}o^A &=& (\bar{\xi}_{0'})^{A'}\nabla_{AA'}(\xi_0)^A =\epsilon^{\Lambda\Delta}\Gamma_{\Delta0'\Lambda0} =\Gamma_{10'00}-\Gamma_{00'10}=\rho-\varepsilon,  \\
	\bar{o}^{A'}\nabla_{AA'}\iota^A &=& (\bar{\xi}_{0'})^{A'}\nabla_{AA'}(\xi_1)^A =\epsilon^{\Lambda\Delta}\Gamma_{\Delta0'\Lambda1} =\Gamma_{10'01}-\Gamma_{00'11}=\alpha-\pi ,  \\
	\bar{\iota}^{A'}\nabla_{AA'}o^A &=& (\bar{\xi}_{1'})^{A'}\nabla_{AA'}(\xi_0)^A =\epsilon^{\Lambda\Delta}\Gamma_{\Delta1'\Lambda0} =\Gamma_{11'00}-\Gamma_{01'10}=\tau-\beta ,  \\
	\bar{\iota}^{A'}\nabla_{AA'}\iota^A &=& (\bar{\xi}_{1'})^{A'}\nabla_{AA'}(\xi_1)^A =\epsilon^{\Lambda\Delta}\Gamma_{\Delta1'\Lambda1} =\Gamma_{11'01}-\Gamma_{01'11}=\gamma-\mu . 
\end{IEEEeqnarray*} 
Consequently, the Weyl equation becomes~\cite{Zhang:2003vb} 
\begin{IEEEeqnarray}{rCl}
	(D+\rho-\epsilon)\psi^1 + (\bar{\delta}+\alpha-\pi)\psi^2 &=& 0 ,  \IEEEyesnumber \IEEEyessubnumber \label{eq:diraceqDpsi1}\\ 
	(\delta+\tau-\beta)\psi^1 + (\Delta+\gamma-\mu)\psi^2 &=& 0 . \IEEEyessubnumber \label{eq:diraceqdeltapsi1}
\end{IEEEeqnarray} 

\section{The spin-weighted spherical harmonics}
\label{sec:swsh}
The spin-weighted spherical harmonics can be defined explicitly as~\cite{Goldberg:1966uu,Penrose:1985bww}
\begin{IEEEeqnarray}{rCl}
	{}_s Y_{lm}(\theta,\varphi) &=& \frac{(-1)^{l+m-s}}{(l-s)!}\ee^{\ii m\varphi} \sqrt{\frac{(l+m)!(l-m)!(2l+1)}{4\pi(l+s)!(l-s)!}} \sin^{2l}\left(\frac{\theta}{2}\right)  \nonumber \\ 
								&&\times \sum_{r=0}^{l-s} (-1)^r \binom{l-s}{r} \binom{l+s}{r+s-m} \cot^{2r+s-m}\left(\frac{\theta}{2}\right). \label{eq:sYlmexplicit}
\end{IEEEeqnarray} 
Such a definition reduces to the ordinary spherical harmonics when $s=0$ and $l$ is a positive integer, namely ${}_0Y_{lm}=Y_{lm}$. In addition, they satisfy the following identities
\begin{IEEEeqnarray}{rCl}
	\eth {}_s Y_{lm} &=& -\left(\partial_\theta+\frac{\ii}{\sin\theta}\partial_\varphi-s\cot\theta\right){}_s Y_{lm}=\sqrt{l(l+1)-s(s+1)}{}_{s+1}Y_{lm} , \label{eq:ethsYlm} \\ 
	\bar{\eth} {}_s Y_{lm} &=& -\left(\partial_\theta-\frac{\ii}{\sin\theta}\partial_\varphi+s\cot\theta\right) {}_s Y_{lm}=-\sqrt{l(l+1)-s(s-1)}{}_{s-1}Y_{lm}  . \label{eq:ethbsYlm} 
\end{IEEEeqnarray}

\section{The static fermionic response of the extremal \acrshort{rn} black hole}
\label{sec:extremalrn}
For the extremal \acrshort{rn} black hole with $Q=M$, the blackening factor $f(r)$ becomes a total square 
\begin{IEEEeqnarray}{rCl}
    f(r)&=&\left(\frac{r-M}{r}\right)^2  .\label{eq:frnextremal}
\end{IEEEeqnarray}
Consequently, the radial Eqs.~\eqref{eq:diraceqradialRm} and~\eqref{eq:diraceqradialRp} reduce to 
\begin{IEEEeqnarray}{rCl}
    \left[-2\ii \omega r^2+(r-M)^2\frac{\dd}{\dd r}\right]R_-(r)&=&-\lambda (r-M)R_+(r) ,\\ 
   (r-M)\frac{\dd}{\dd r}R_+(r)&=&-\lambda R_-(r) .
\end{IEEEeqnarray}
Similar to the non-extremal case, we can eliminate $R_+(r)$ and obtain the second-order equation for $R_-(r)$. But the dimensionless variable $z$ introduced in \eqref{eq:defz} is ill defined for the extremal case, so we like to introduce a new dimensionless variable
\begin{IEEEeqnarray*}{rCl}
    \tilde{z}&=&\frac{r-M}{M},
\end{IEEEeqnarray*}
in terms of which the second-order equation for $R_-$ can be written as 
\begin{IEEEeqnarray}{rCl}
   \tilde{z}^3 \frac{\dd^2R_-}{\dd \tilde{z}^2}-\ii \tilde{z} (\ii \tilde{z}+2M\omega+4M\omega \tilde{z}+2M\omega \tilde{z}^2)\frac{\dd R_-}{\dd\tilde{z}}-(\lambda^2\tilde{z}-2\ii\omega M+2\ii M\omega \tilde{z}^2 )R_-&=&0 .\nonumber \\
\end{IEEEeqnarray}
In the static case with $\omega=0$, the above equation becomes
\begin{IEEEeqnarray}{rCl}
   \tilde{z}^2\frac{\dd^2 R_-}{\dd \tilde{z}^2}+\tilde{z}\frac{\dd R_-}{\dd \tilde{z}}-\lambda^2 R_-&=&0 , \label{eq:Rmztpp}
\end{IEEEeqnarray}
whose general solution has the following form 
\begin{IEEEeqnarray}{rCl}
   R_-(\tilde{z}) &=& c_1 \tilde{z}^{-\lambda}+c_2 \tilde{z}^{\lambda} . \label{eq:Rmztppsol}
\end{IEEEeqnarray}
The regularity on the horizon requires us to set $c_1=0$.  For simplicity but without loss of generality, we can set $c_2=1$ as well. Accordingly, our regular solution behaves as follows
\begin{IEEEeqnarray}{rCl}
   R_-(\tilde{z}) &=&  \tilde{z}^{\lambda} \underset{r\to\infty}{\to} \frac{r^{l+\frac{1}{2}}}{M^{l+\frac{1}{2}}} \label{eq:Rmztppsolreg}
\end{IEEEeqnarray}
at infinity,
where we have used the fact that $\lambda=l+\frac{1}{2}$. 
The absence of the $\left(\frac{r_+}{r}\right)^{2l+1}$ term in the large $r$ expansion indicates that $\mathcal{F}_{-\frac{1}{2}lm}$ vanishes in the extremal case. The same behavior applies to  $\mathcal{F}_{\frac{1}{2}lm}$.

\printglossary

\input{./main.bbl}

\end{document}

%% file: main.bbl
\providecommand{\href}[2]{#2}\begingroup\raggedright\endgroup